\documentclass[12pt]{JHEP3}

\usepackage{epsfig}

\newcommand{\bmat}{\left(\begin{array}}
\newcommand{\emat}{\end{array}\right)}

\def\yzero{\smash{\hbox{$y\kern-4pt\raise1pt\hbox{${}^\circ$}$}}}

\def\beq{\begin{equation}}
\def\eeq{\end{equation}}
\def\beqa{\begin{eqnarray}}
\def\eeqa{\end{eqnarray}}

\def\-{\hphantom{-}}

\def\s2{\frac{1}{2}}

\def\beq{\begin{equation}}
\def\eeq{\end{equation}}
\def\beqa{\begin{eqnarray}}
\def\eeqa{\end{eqnarray}}

\def\IF{\relax{\rm I\kern-.18em F}}
\def\II{\relax{\rm I\kern-.18em I}}
\def\IP{\relax{\rm I\kern-.18em P}}
\def\IC{\relax{\rm I\kern-.48em C}}
\def\IR{\relax{\rm I\kern-.18em R}}

\def\cn{{\cal N}}

\def\cp{{\cal P}}

\def\Dsl{\,\raise.15ex\hbox{/}\mkern-13.5mu D} 
\def \one{\relax{\rm 1\kern-.26em I}}

 \def\cp#1{\relax\ifmmode {\IP\kern-2pt{}_{#1}}\else $\IP\kern-2pt{}_{#1}$\=fi}

\catcode`\@=11
\newdimen\@rotdimen
\newbox\@rotbox

\def\@vspec#1{\special{ps:#1}}
\def\@rotstart#1{\@vspec{gsave currentpoint currentpoint translate
   #1 neg exch neg exch translate}}
\def\@rotfinish{\@vspec{currentpoint grestore moveto}}
\def\@rotr#1{\@rotdimen=\ht#1\advance\@rotdimen by\dp#1%
   \hbox to\@rotdimen{\hskip\ht#1\vbox to\wd#1{\@rotstart{90 rotate}%
   \box#1\vss}\hss}\@rotfinish}
\def\@rotl#1{\@rotdimen=\ht#1\advance\@rotdimen by\dp#1%
   \hbox to\@rotdimen{\vbox to\wd#1{\vskip\wd#1\@rotstart{270 rotate}%
   \box#1\vss}\hss}\@rotfinish}%
\def\@rotu#1{\@rotdimen=\ht#1\advance\@rotdimen by\dp#1%
   \hbox to\wd#1{\hskip\wd#1\vbox to\@rotdimen{\vskip\@rotdimen
   \@rotstart{-1 dup scale}\box#1\vss}\hss}\@rotfinish}%
\def\@rotf#1{\hbox to\wd#1{\hskip\wd#1\@rotstart{-1 1 scale}%
   \box#1\hss}\@rotfinish}%
\def\rotate{\@ifnextchar[{\@rotate}{\@rotate[l]}}
\def\@rotate[#1]#2{\setbox\@rotbox=\hbox{#2}\@nameuse{@rot#1}\@rotbox}

\catcode`\@=12

\title{de Sitter String Vacua \\
from Supersymmetric D-terms}

\author{C.P. Burgess,$^1$ R. Kallosh$^2$ and F. Quevedo$^3$
\\
$^1$Physics Department McGill University\\
 3600 University Street\\
 Montr\'eal, Quebec, Canada, H3A 2T8.\\
\\
$^2$Department of Physics, Stanford University,
\\ Stanford, CA 94305-4060, USA.\\
\\
$^ 3$DAMTP, Centre for Mathematical Sciences\\
               University of Cambridge,\\
               Cambridge CB3 0WA UK.}

\abstract{We propose a new mechanism for obtaining de Sitter vacua in type IIB string
theory compactified on (orientifolded) Calabi-Yau manifolds
similar to those recently studied by Kachru, Kallosh, Linde and Trivedi
(KKLT).
dS vacuum appears in KKLT model after uplifting an AdS vacuum by adding an anti-D3-brane, which explicitly breaks supersymmetry. We 
accomplish the same goal by adding fluxes of gauge fields within the
D7-branes, which induce a D-term potential in the effective 4D action. In
this way we obtain dS space as a spontaneously broken vacuum from
a purely supersymmetric 4D action.
We argue that our approach can be directly extended to heterotic string
vacua, with the dilaton potential obtained from a combination of gaugino
condensation and the D-terms generated by anomalous $U(1)$ gauge groups.}

\preprint{NSF-KITP-03-81, DAMTP-2003-90, McGill-03/21, SU-ITP-03/26  }

\begin{document}

\makeatletter \@addtoreset{equation}{section} \makeatother
\renewcommand{\theequation}{\thesection.\arabic{equation}}







\setcounter{page}{1} \pagestyle{plain}
\renewcommand{\thefootnote}{\arabic{footnote}}
\setcounter{footnote}{0}


\section{Introduction}
Recently Kachru, Kallosh, Linde and Trivedi (KKLT) \cite{kklt} found the
first explicit realization of 4D de Sitter space as a solution to the
low-energy equations of string theory. This is a significant
achievement
given the importance that de Sitter space has acquired from the recent
data on the acceleration of the universe and also for its close relation
with the inflationary scenario.

The proposal of KKLT combines several mechanisms that lift the vacuum
degeneracy of supersymmetric string models. First they introduce
background fluxes for NS and RR forms to fix all of the complex-structure
moduli of a Calabi-Yau compactification \cite{fluxes, gkp, gvw, kpv, bp}.
Second, they focus on models having only one K\"ahler modulus, and fix
this remaining modulus using nonperturbative effects combined with the
remnant superpotential produced by the fluxes, typically leading to a
supersymmetric 4D anti-de Sitter vacuum. Finally, the addition of an
anti-D3-brane provides an extra source of positive potential energy which
depends on the K\"ahler modulus, and lifts the minimum to a 4D de Sitter
vacuum.

All steps except the last one can be understood within the context of an
effective 4D supergravity, but since the addition of the anti-D3-brane
breaks supersymmetry,
the effective 4D theory cannot be put into the standard 4D supergravity
form.\footnote{It may be possible to write the 4D action in a manifestly
supersymmetric way if certain massive fields are `integrated in' to act
as the superpartners of fields appearing in the non-supersymmetric 4D
effective action.} This kind of explicit breaking of supersymmetry
considerably complicates the analysis of the low-energy theory, because
the loss of supersymmetry removes much of the theoretical control over
the types of interactions which can be induced.
This motivates searching for ways to obtain de Sitter space in string
theory, but within a framework described by a fully supersymmetric action.

In this note we do so by modifying the last step of the KKLT
stabilization process. Instead of adding an anti-brane, we turn on fluxes
for yet another field: the gauge fields living inside the D7 branes. This
extra degree of freedom is already contained within the original scenario
and does not require adding sources such as anti-branes which are
difficult to interpret as an F- or a D-term of a potential of a
supersymmetric theory. Nevertheless, the addition of these fluxes has the
similar effect of adding a positive term to the scalar potential, which
can be interpreted as a D-term of the effective 4D supersymmetric action.
It is the competition of this D-term potential with the superpotential
contribution which breaks supersymmetry spontaneously and leads to a 4D
de Sitter geometry.

Another method for replacing the antibrane in the KKLT construction is to
obtain models whose potential at the no scale level admits a minimum with
$DW\ne 0$ so that the potential has an additional term similar to the
antibrane contribution to the potential \cite{Eva}.

\section{Modulus Stabilization}
Type IIB strings have RR and NS-NS antisymmetric 3-form field strengths,
$H_3$ and $F_3$ respectively, which can  wrap on 3-cycles of the
compactification manifold, leading to quantized background fluxes 
\beqa
 \frac{1}{4\pi^2 \alpha'} \int_A F_3\ = M \, , \qquad 
 \frac{1}{4\pi^2 \alpha'} \int_B H_3\ = -K\,,
\eeqa
where $K$ and $M$ are arbitrary integers and $A$ and $B$ label the
different 3-cycles of the Calabi-Yau manifold.

Fluxes have proven to be very efficient at fixing many of the string
moduli, including the axion-dilaton field of type IIB theory, $S= e^\phi
+ia$ \cite{fluxes, gkp}. A very general analysis of orientifold models of
type IIB, or its equivalent realization in terms of $F$-theory, has been
done in \cite{gkp}. In the $F$ theory approach, the geometrical picture
corresponds to an elliptically fibered four-fold Calabi-Yau space $Z$
with base space ${\cal M}$ and the elliptic fiber corresponding to the
axion-dilaton field $S$.

An important consistency condition coming from tadpole cancellation
implies a relationship between the charges of D-branes, O-planes, and
fluxes, which can be written as follows:
\beq \label{tadpole}
 N_{D3}-N_{\bar D3}+ N_{flux}= \frac{\chi(Z)}{24}\,.
\eeq
Here the left-hand side counts the number of D3-branes and anti-branes,
as well as the flux contribution to the RR charge:
\beq
 N_{flux}= \frac{1}{2\kappa_{10}^2 T_3} \int_M H_3\wedge
 F_3\,.
\eeq
The right-hand side of (\ref{tadpole}) refers to the Euler number of the
four-fold manifold, $Z$, or in terms of orientifolds of type IIB, to the
contribution of the D3-brane charge due to orientifold planes and
D7-branes.  Here $\kappa_{10}$ is the string scale in 10D and
$T_3$ is the tension of the D3 branes.

{}From the point of view of the effective 4D theory, the fluxes generate a
superpotential of the Gukov-Vafa-Witten form \cite{gvw}:
\beq
 W\ =\ \int_M G_3\wedge \Omega\,,
\eeq
where $G_3 = F_3 - iS \, H_3$ with $S$ the dilaton field and $\Omega$ is
the unique $(3,0)$ form of  the corresponding Calabi-Yau space. An
important feature of this superpotential is that it is independent of all
of the compact space's K\"ahler moduli,  so none of these moduli are
fixed by these kinds of fluxes.

The simplest models have the fewest possible number of K\"ahler moduli: one. It
defines the overall breathing mode of the underlying Calabi-Yau space.
Denoting the 10D metric by $ds_{10}^2 = L^{-6} \, ds_4^2 + L^2 \,
ds_6^2$, this K\"ahler mode can be written $T=X+iY$, where $X = L^4/g_s$
and $Y$ is an axion field coming from the
RR 4-form. ($T=i\rho$ in the conventions of \cite{gkp,kklt}.) Since this
modulus does not appear in $W$, it is not fixed by the fluxes, and the
corresponding flat direction can be understood within the effective 4D
theory because it takes the no-scale form \cite{noscale}. That is, the
low-energy K\"ahler potential is:
\beq
 K\ =\ \tilde K (\varphi_i, \varphi_i^*) -3 \log
 \left( T+ T^*\right)\,,
\eeq
with $\tilde K$ the K\"ahler potential for all of the other fields,
$\varphi_i$, except for $T$. The corresponding F-term  
 potential then takes the form
\beqa
 V_{F} &=& e^K\left( K^{I\bar J} D_IW \, \overline{D_{J} W} - 3 \, |W|^2
 \right)\nonumber \\
 &=& e^K\left(K^{i\bar j} D_iW \overline{D_{j} W}\right)\,,
\eeqa
where $i,j$ label the fields $\varphi^i$, while the labels $I,J$ include
both the $\varphi^i$ and $T$. Here $K^{i\bar j}$ is the inverse of the
K\"ahler metric, $K_{i \bar j} =
\partial_i \partial_{\bar j} K$, and $D_i W =\partial_i W + W\partial_i
K$ is the superpotential's K\"ahler covariant derivative. The second
equality follows because the $T$-dependence of the K\"ahler potential is
such that the contribution $K^{T\bar T} |D_T W|^2$ precisely cancels the
term $-3|W|^2$. This  is a special property of no-scale models
\cite{noscale}. Being positive definite, the global minimum of this
potential lies at zero, with all fields except for $T$ fixed by the
conditions $D_i W=0$. This minimum is supersymmetric if $D_TW = W=0$, and
not supersymmetric otherwise.

In order to fix $T$ KKLT proceed as follows. First, they choose fluxes to
obtain a vacuum in which supersymmetry is broken by the $T$ field,
because  $W = W_0\neq 0$. Then they consider a nonperturbative
superpotential, either generated by Euclidean D3-branes or by gaugino
condensation within a non-abelian gauge sector of $N$ wrapped  D7-branes.
Since the gauge coupling for such a gauge group is ${8\pi^2}/{g_{YM}^2} =
2\pi {L^4}/{g_s} = 2\pi X$, well-established arguments imply a
non-perturbative superpotential of the form $W_{np}= A e^{-aT}$
\cite{gauginocondensation}, for appropriate constants $A$ and $a$.

Combining the two sources of superpotentials
\beq
 W= W_0 + A e^{-aT},
\eeq
gives an effective scalar potential for the field $T = X+iY$ of the form:
\beq
 V_F  =  \frac1{8 X^3} \left\{ \frac13|2X W^\prime
 - 3 W|^2 - 3|W|^2 \right \}\,, \label{spot}
\eeq
having a non-trivial minimum at finite $T$ as well as the standard
runaway behaviour towards infinity. The non-trivial minimum corresponds
to negative cosmological constant, and so gives rise to a supersymmetric
AdS vacuum.

This is reminiscent of the standard situation in heterotic string theory
regarding the potential for the dilaton field $S$ coming from the
racetrack scenario, after having fixed the field $T$.
More general superpotentials have been considered in
\cite{egq}, in which the superpotential takes the form:
\beq
 W= \sum_i A_i e^{-a_iT},
\eeq
where the sum can be finite or infinite. The finite case is the standard
racetrack scenario \cite{krasnikov}. The structure of the scalar
potential is such that it has one or many anti-de Sitter minima, all of
which are supersymmetric.

\section{de Sitter Vacua}
\begin{figure}[h]
  \def\epsfsize#1#2{0.6#1}
  \centerline{\epsfbox{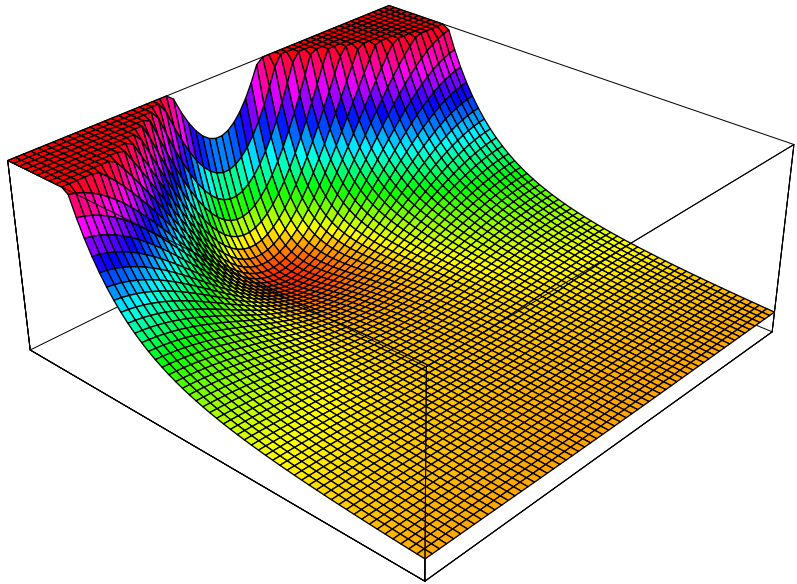}}
  \caption{\it The scalar potential considered in \cite{kklt}
  with a de Sitter minimum.}
  \label{figure:kkltpotential}
\end{figure}

In order to obtain de Sitter vacua, KKLT add anti-D3 branes to the above
construction, while still satisfying the tadpole condition
(\ref{tadpole}). Semiclassically, this has the effect of adding an extra
non-supersymmetric term (neither F- nor D-term) to the scalar potential
of the form:
\beq
 V \ =\ V_F + \frac{k}{X^2} \,,
\eeq
with the constant $k=2 a_0^4 T_3/g_s^4$ parameterizing the lack of
supersymmetry of the potential. 
Here $a_0$ is the warp factor at the
location of the anti-D3 branes and $T_3$ is the anti-brane tension. The
net effect of this addition to the potential is that, for suitable values
of $k$, the original AdS minimum gets lifted to a dS one with broken
supersymmetry. Fig.~(\ref{figure:kkltpotential}) gives a sample plot of
the potential obtained in this way.

We 
will depart from the KKLT proposal and choose not to introduce
anti-D3-branes, and instead consider the possibility of turning on
non-trivial fluxes for the gauge fields living on the D7 branes. From the
linear D7 action, these fluxes will naturally induce a term in the 4D
effective action of the form:
\beq\label{fluxd7}
T_7 \int_\Gamma d^4y \sqrt{g_8}\, F^{mn} F_{mn} = \frac{2 \pi \, E^2}{X ^3} \,,
\eeq
where $\Gamma$ is the 4-cycle about which the D7 branes wrap, $T_7$
their tension, $E$ is a
measure of the strength of the flux and the conventional factor of $2\pi$
is for later convenience. The $X$ dependence can be inferred by a simple
scaling analysis, using $\sqrt{g_8} \, g^{mn} g^{pq} \propto L^{-12}$.

The contribution of this term by itself is similar to the
contribution induced by the introduction of anti-D3 branes. The only
difference is the power of $X$ in the FI term.\footnote{The power of $X$ in the D3-brane term was $-3$ in
ref.~\cite{kklt}, and this was corrected in ref.~\cite{kklmmt} to be $-2$ due
to the appearance of an extra power of $X$ in the warp factor.}

The supersymmetry of this type of term from the 4D point of view has been
studied in the past elsewhere, and was identified as arising as a
field-dependent Fayet-Iliopoulos (FI) D-term in the $\cn=1$
supersymmetric effective action \cite{witten, csu}. In general, type A
branes --- corresponding to D-branes wrapping 3-cycles ---
couple to complex-structure moduli as FI terms, while branes of type B
--- corresponding to D-branes wrapping around cycles of even
dimensionality and with magnetic fluxes inside --- couple to the
K\"ahler-structure moduli as FI terms \cite{witten}. It is the second of
these which is of interest for the present purposes.\footnote{ We thank
A. Uranga for a conversation on this point.}

The full 4D-supersymmetric contribution to the scalar potential
corresponding to eq.~(\ref{fluxd7}) therefore takes the form:
\beq \label{VDexpr}
 V_D\ =\ {g^2_{YM} \over 2} \, D^2 = \frac{2\pi}{X}
 \left(\frac{E}{X} +\sum{q_I |Q_I|^2}\right)^2 \, ,
\eeq
where we again use $4\pi/g^2_{YM} = X$. The $Q_I$ represent any matter
fields which are charged --- with charges $q_I$ --- under the $U(1)$
gauge group for which the fluxes provide a FI factor, such as can arise
from D3-D7 strings. Eq.~(\ref{VDexpr}) assumes for simplicity a minimal
K\"ahler potential for $Q_I$ of the form $|Q_I|^2$, although this is not
crucial for the later discussion.

If the additional fields are minimized at $Q_I = 0$, then the total
potential
\beq
 V\ = \ V_F + V_D \,
\eeq
behaves very similarly to the KKLT potential, with $V_D=2\pi E^2/X^3$
 playing
the role played by their anti-brane term. The net result for $X$ is
therefore qualitatively the same as for KKLT, but with the difference
that we use only a fully supersymmetric 4D effective action. Whether the
potential has de Sitter or anti-de Sitter minima depends on the values of
the flux parameter $E$. If $E$ is too small, the minimum remains anti de
Sitter, but if it belongs to a range of slightly larger values, then there
is a de Sitter minimum. Beyond a particular critical value, the runaway
behaviour of the pure FI potential dominates and the local minimum
disappears, leaving only the runaway minimum at $X \to \infty$.

It is important for this argument that any fields $Q_I$ indeed get
minimized at $Q_I = 0$, since the stabilization of the de Sitter vacuum
cannot occur along the lines described here if the $Q_I$ instead adjust
to ensure $V_D = 0$. Whether this happens for a particular string vacuum
is model dependent, but we pause here to describe what can be said on
reasonably general grounds.

If we would consider a system of D7 brane with fluxes and a D3 brane at
some distance as in \cite{Herdeiro:2001zb},
 we would have to take into account D3-D7 strings corresponding to positive and negative
  charged fields. They are stretched between separated D7 and D3 branes
   and their mass  depends on the distance between branes. When D3 brane
dissolves onto D7 brane with fluxes,  the D-flatness condition  $V_D = 0$
is realized.  Here, however, it is more appropriate to consider a single
D7 brane with fluxes. If the D3-D7 separation is very large, then the
D3-D7 open strings can be very massive, perhaps leading to no charged
fields $Q_I$ in the effective 4D theory. A similar situation may occur if
there are only fluxes but no D3 branes.

Note also that due to the presence of a flux on a D7 brane, a D5 brane
charge may be generated, in general, via a Chern-Simons term $\int
F_2\wedge C_6$. For our purpose we may consider either the situations
where $\int F_2$ over a 2-cycle inside of a 4-cycle,  which D7 brane
wraps, vanishes, or a situation when somewhere far  there is an
anti-D5-brane so that the total D5 charge vanishes. In orientifold
models
the orientifold action tends to change the sign of the gauge fields inside the 
D7 branes and then odd powers of $F_2$ do not contribute to tadpoles,
in particular, D5 brane tadpoles are absent \footnote{We thank  R. Blumenhagen, S. Kachru
 and R. Rabad\'an for the useful conversations on these issues.}.
A D3 brane tadpole, generated by $\int C_4\wedge F_2^2$ only contributes
to an extra term to equation (2.2) that can always be taken care of by
the choice of the original flux contribution.

Second, it can happen that even if the $Q_I$'s appear in the 4D effective
theory, in some circumstances the scalar potential is minimized at $Q_I =
0$. Even if $W$ is independent of $Q$, $V_F$ typically depends on $Q$
because the K\"ahler derivative is $|D_I W|^2 = |K_I W|^2 = |Q_I W|^2$.
As is clear from these expressions, $V_F$ is often minimized by $Q_I =
0$. Since the $Q$-dependent terms in $V_F$ are proportional to $X^{-3}$,
while those in $V_D$ vary like $X^{-1}$, for large enough $X$ it is preferable for
$Q_I$ to adjust to minimize $V_D$ rather than $V_F$. Conversely, for
smaller $X$ the relative importance of $V_F$ for the minimization of $Q_I$
increases. An extreme case arises if only fields having a charge with the
same sign as $E$ appear nontrivially in $W$, in which case there is no
value for these fields for which $V_D = 0$.
\footnote{In general we do not need to have the matter fields to
  vanish. As long as the D-term part of the potential is non-zero at the
  minimum, it may do the job of lifting the vacuum  to de Sitter
  space. In ref.~\cite{bddp}, a particular example was worked out where
  it was shown that the combination of gaugino condensation and FI
  D-terms
implies supersymmetry breaking in the sense that the D-term was not
  zero at the minimum.}

This last situation is analogous to the hybrid D-term inflation models
\cite{Binetruy:1996xj}.  Some of the superfields, which are left out of the
superpotential, have positive charge only and therefore, for positive FI
terms they have positive mass squared and vanish at the minimum of the potential
\footnote{For the  earlier work on single field inflation models with
anomalous FI terms and a charged   inflaton see \cite{Casas:1988pa}.}. The
negatively charged fields which would be able to cancel the contribution
from fluxes should be absent in this model since they would have a
negative mass squared, unless they appear in the particular form (as in D-term
inflation models) in the superpotential. However, the original D-term
inflation model \cite{Binetruy:1996xj} as well as the brane construction
\cite{Herdeiro:2001zb} realizing the D-term inflation in string theory, do not include the
4D space-dependent volume moduli and need to be properly generalized
before the complete picture can be clarified. For our specific purpose of
describing a possibility of a de Sitter minimum in presence of fluxes on
D7 branes, the absence of negatively charged fields seems to be a 
sufficient
condition for the absence of tachyons as well as for  vanishing
$Q_I$-fields at the minimum.

Although our starting point was the assumption of nonvanishing
electromagnetic flux,  our results only depend on the
existence of a FI term, regardless of its origin. The only requirement is
that the FI term depends on the (real part of the) $T$-field as an
appropriately negative power of $X$. Let us pause here to understand this
point better.

A sufficient condition for the existence of the FI term is the existence
of a Green-Schwarz type of coupling of the form $B \wedge F$ between a
$U(1)$ gauge field and the antisymmetric tensor, $B$, dual to the
imaginary part of $T$.\footnote{Recall that $Y$ is dual to the two-index
field $B_{\mu\nu}$ defined by $C_{\mu\nu ab}= B_{\mu\nu} J_{ab}$ where
$C_{MNPQ}$ is the original 4-index field of type IIB string theory and
$J_{ab}$ is the K\"ahler form of the Calabi-Yau \cite{gkp}.} This kind of
coupling can arise, for instance, from the Chern-Simons coupling of the
D7 brane, which takes the form:
\beq
 \int_{{\cal M}_8} C\wedge \left(Tr\ e^{\frac{iF}{2\pi}}\right) \,.
\eeq
Expanding the exponential produces the coupling $C_{\mu\nu
mn}F_{\rho\sigma}F_{pq} \epsilon^{\mu \cdots q}$, where, as usual, Greek
indices denote ordinary spacetime and Latin indices run over the
Calabi-Yau dimensions. This precisely gives rise to the $B\wedge F$ term
in the presence of nonvanishing background fluxes.

The necessity for a FI term follows from supersymmetry and gauge
invariance, because a $B \wedge F$ term becomes $A_\mu\partial^\mu Y$
when expressed in terms of the axion, $Y$, which is dual to $B$. But gauge
invariance then requires that $Y$ and $A_\mu$ must always appear together in
the combination $\partial_\mu Y - q A_\mu$ for some constant $q$, and
supersymmetry then implies that the K\"ahler potential can depend on $T$
only through the combination $T+T^* +c V$
\footnote{Notice that, since $T$ transforms under the $U(1)$ interactions,
 a simple exponential
of $T$ in the non-perturbative superpotential is not invariant under the
corresponding $U(1)$ and needs to be compensated by the dependence of the
superpotential on other fields. This, being  model dependent, is beyond
the scope of the present note. Nevertheless, it would be interesting to
fully study  a particular example with the  matter field dependence under
control.}. Here $V$ is the gauge superfield including $A_\mu$ and $c$ is
a constant (proportional to $E$).
 Expansion of the K\"ahler function, $K(T + T^* + cV)$,
into components then induces a FI D-term proportional to $\partial
K/\partial V\vert_{V=0}$. Notice that whenever $K$ has the no-scale form,
$K = - 3 \ln(T + T^* + cV)$, the resulting auxiliary field D is
proportional to $E/X$, as required by expression (\ref{VDexpr}) for $V_D$
above, so that the potential has a standard form ${1\over 2} g^2 D^2$.

Clearly our construction should generalize to any string vacuum for which
such a $B \wedge F$ coupling appears. A broad (but not exhaustive) class
of vacua for which it does consists of those for which there is an
anomalous $U(1)$ in the low-energy theory, with $Y$ participating in the
Green-Schwarz anomaly cancellation.

\section{Heterotic String Vacua}
Most of the past work on moduli potentials was done for heterotic string
vacua, and potentials very similar to the ones appearing in the type IIB
case were actually computed in that approach. A fixing of all string
moduli has not yet been achieved for heterotic vacua, although work in
this direction is now in progress by several groups \cite{Shamit}. In all
cases when particular moduli were assumed fixed, the potentials for some of
the remaining fields were sometimes calculable and were always found to
be minimized for 4D anti-de Sitter vacua. One might wonder if there is a
way to lift those vacua to the de Sitter case as well. At first sight,
this seems difficult if we follow the KKLT scenario, since there is no
analogue of the anti-D3-brane in the heterotic case. The same objection
does not apply to the analysis as presented in the previous section,
however, which is straightforward to generalize to the case of heterotic
strings.

Nonperturbative superpotentials were studied in order to fix the `model
independent' moduli fields, $T$ and $S$, of the heterotic string using
the standard tree-level K\"ahler potential \cite{truncation}:
\beq
 K\ = \ -\log\left(S+S^*\right)\ -3\log\left(T+T^*\right) \,.
\eeq
Since $S$ is in this case the gauge coupling, the nonperturbative
superpotential takes the `racetrack' form:
\beq
 W(S,T)\ =  \sum_i A_i\left(T\right) e^{-a_i S} \,,
\eeq
where the $S$ dependence comes from gaugino condensation for different
gauge groups, and the $T$ dependence can be obtained either from
threshold corrections to the gauge coupling or by just requiring
invariance under $T$-duality of the heterotic string \cite{filq,
cfilq,dccm}. We do not here consider any details regarding the functional
dependence on $T$, but just content ourselves with pointing out that once
$T$ is fixed we are left with a potential, $V_F$, for $S$ having the same
form as the ones considered in the previous section for $T$. The generic
potential of this form fixes the field $S$ at a supersymmetric or
non-supersymmetric minimum with a negative (anti-de Sitter) cosmological
constant.

It is also true that in generic heterotic models there is usually at
least one anomalous $U(1)$ group. The anomaly is cancelled by a
four-dimensional version of the Green-Schwarz mechanism, using the NS-NS
antisymmetric field $B_{\mu\nu}$. Writing the dual of this field as
$\partial_\mu a = \epsilon_{\mu\nu\rho\sigma}\partial^\nu
B^{\rho\sigma}$, the couplings required by anomaly cancellation are $a
F_{\mu\nu}{\tilde F}^{\mu\nu}$ and $B\wedge F$. Recall that $a$ is the axion field
that makes up the imaginary part of the field $S$ and $F_{\mu\nu}$ is the
field strength of the anomalous $U(1)$ gauge group.

As is argued above (and is well known \cite{dsw}), the existence of the
$B \wedge F$ coupling implies (from gauge-invariance and supersymmetry)
the existence of a FI term for the corresponding $U(1)$. We are again led
to a D-term potential of the form:
\beq
 V_D\ =\
 \frac{1}{S+S^*}\left(\frac{\hbox{Tr}\ q}{48\pi^2}\frac{1}{
 (S+ S^*)^2}+\sum{q_I |Q_I| ^2}\right)^2 \,,
\eeq
where the first term in the bracket is the FI term.

We see that the situation is completely analogous to our previous
discussion for the type IIB string, with the roles of $S$ and $T$
interchanged. After considering the full scalar potential $V= V_F +
V_D$\footnote{For discussions on the combination of F-terms and D-terms
induced by anomalous $U(1)$'s see for instance \cite{bddp}.} we can have
a local minimum with vanishing charged fields $Q_I$ and with $S$ fixed at
a de Sitter minimum.

Whether a de Sitter minimum is really achieved is a model-dependent
question. The important quantity to follow is the coefficient $\hbox{Tr}\
q$. If there are many charged fields giving rise to a relatively large
`anomaly' $\hbox{Tr}\ q$, the D-term can dominate and lift the anti-de
Sitter minimum to a de Sitter one. Otherwise the minimum remains anti-de
Sitter. If $\hbox{Tr}\ q$ is too large, then the FI term dominates the
potential and may completely remove the minimum at finite $S$.

\section{Conclusions}
In this paper we generalize the KKLT construction \cite{kklt} of de
Sitter string theory vacua for which all of the moduli are fixed. The
main difference between our approach and the previous one is the
derivation from string theory of a metastable de Sitter space via a
spontaneous supersymmetry breaking. It is achieved in the framework of
the effective 4D supersymmetric effective action, through the interplay
of F- and D-terms in the scalar potential. The virtue of keeping the
low-energy supersymmetry manifest is the extra control this gives over
the form of the low-energy Lagrangian, which is more difficult for the
anti-D3-branes used by KKLT.

There are several issues in our proposal which may need to be examined in
more detail for particular examples. Central among these is the question
of finding systems for which the matter fields $Q_I$ do not cancel the
contribution from fluxes to $V_D$. A second important issue is to
understand whether new moduli arise when fluxes are turned on the D7
brane. Finally, it would be worth understanding in more detail the
relation of our flux
proposal 
 to the D3/D7 system of the type described in
\cite{Herdeiro:2001zb}. Notice that the D3/D7 system is analogous to
Seiberg-Witten non-commutative theories \cite{Seiberg:1999vs} and the
D0/D4 system of branes where the FI terms are related to fluxes on the D4
brane.

Even though our technique is described in a supersymmetric context, our
result is very similar to the one obtained by KKLT by adding an
anti-D3-brane.\footnote{ An anti-D3 brane can dissolve
into an anti self-dual flux in a D7 brane, so this class of constructions
may be related to the original KKLT construction as a different branch of
the anti-D3 configuration space in absence of D5 brane charge. We thank
S. Kachru, J. Maldacena, E. Silverstein and A. Uranga for the discussion
of this possibility.}

In contrast with the previous scenario, ours appears to immediately
generalize to the heterotic string cases that have been studied in the
past. It would be interesting to approach the possibility of obtaining
D-brane inflation
 \cite{dbraneinflation,Herdeiro:2001zb}
 from our scenario,
along the lines of \cite{kklmmt}.

We believe that our results set the existence of de Sitter
space in string theory on firmer grounds, and are likely to be among the
first of many new constructions of de Sitter vacua in string theory.

\section{Acknowledgements}
We thank the organizers of the KITP workshop on string cosmology for
providing the perfect environment to start this work. We would like to
thank  P. Binetruy, J. Blanco-Pilado, R. Blumenhagen, 
 R. Brustein, C. Escoda, M.
G\'omez-Reino, S. Kachru, S. Kovacs, P. Kumar, A. Linde, J. Maldacena,
V.S. Nemani, R. Rabad\'an,
 E. Silverstein, H. Tye and A. Uranga for many discussions on
these and related subjects. This research was supported in part by the
NSF under grant No. PHY-99-07949. C.B. is also funded by NSERC (Canada),
FCAR (Qu\'ebec) and McGill University.
 The work of R. K.
is supported also by the NSF  grant PHY-0244728. F.Q. is partially funded
by PPARC.

\end{document}